\documentclass[prb,showpacs,preprintnumbers,amsmath,amssymb,floatfix,preprint,superscriptaddress]{revtex4-1}
\usepackage{hyperref}
\usepackage{epsfig}
\usepackage{color}
\begin{document}
\title{The breakdown of superlubricity by driving-induced commensurate dislocations}
\author{A. Benassi}
\email{andrea.benassi@nano.tu-dresden.de}
\affiliation{Institute for Materials Science and Max Bergmann Center of Biomaterials, TU Dresden, 01062 Dresden, Germany}
 \affiliation{Dresden Center for Computational Materials Science (DCCMS), TU Dresden, 01062 Dresden, Germany}

\author{Ming Ma}
\affiliation{School of Chemistry, Tel Aviv University, 69978 Tel Aviv, Israel}

\author{M. Urbakh}
\email{urbakh@post.tau.ac.il}
\affiliation{School of Chemistry, Tel Aviv University, 69978 Tel Aviv, Israel}

\author{A. Vanossi}
\email{vanossi@sissa.it}
\affiliation{CNR-IOM Democritos National Simulation Center, Via Bonomea 265, 34136 Trieste, Italy}
\affiliation{International School for Advanced Studies (SISSA), Via Bonomea 265, 34136 Trieste, Italy}

\begin{abstract}
In the framework of a Frenkel-Kontorova-like model, we address the robustness of the superlubricity phenomenon in an edge-driven system at large scales, 
highlighting the dynamical mechanisms leading to its failure due to the slider elasticity. The results of the numerical simulations perfectly match the length critical size derived 
from a parameter-free analytical model. By considering different driving and commensurability interface configurations, we explore the distinctive nature of the transition 
from superlubric to high-friction sliding states which occurs above the critical size, discovering the occurrence of previously undetected multiple dissipative jumps in the friction force 
as a function of the slider length. These  driving-induced commensurate dislocations in the slider are then characterized in relation to their spatial localization and width, depending on 
the system parameters. Setting the ground to scale superlubricity up, this investigation provides a novel perspective on friction and nanomanipulation experiments and can serve 
as a theoretical basis for designing high-tech devices with specific superlow frictional features.
\end{abstract}

\maketitle
\section{Introduction}
In the emerging field of nanoscale science and technology, understanding both the
statics and the non-equilibrium dynamics of systems with many degrees of freedom in interaction with
some external potential, as is commonly the case in solid state physics, is becoming a crucial
issue. Friction belongs to this category too, because the microscopic asperities of the mating
surfaces may interlock and give rise to intriguing length scale competition mechanisms.

Frequently, despite the basic level of details, simple phenomenological models of friction
\cite{revmodphys,urbakh_review} have revealed the ability of describing the main features of the complex
microscopic dynamics, providing good qualitative agreement with experimental results on nanoscale
tribology, or with more complex simulation data of sliding phenomena. With respect to this, the
application of driven Frenkel-Kontorova-like (FK) systems \cite{Braunbook,vanossiJPCM}, modeling 
the dissipative dynamics of a chain of interacting particles that slide over a rigid substrate potential 
due to the application of an external driving force, has found increasing interest as a possible interpretative key 
of the atomistic processes occurring at the interface of two materials in relative motion.

In particular, the remarkable idea of frictionless sliding connected with interface
incommensurability, a pervasive concept of modern tribology named superlubricity \cite{erdemir}, can be
mathematically drawn in the framework of the FK model \cite{aubry}. When two contacting
crystalline workpieces are incommensurate, due to lattice mismatch or to angular misalignment, the
minimal force required to achieve sliding, i.e. the static friction, should vanish, provided the two
substrates are stiff enough. In such a geometrical configuration, the surface mismatch can prevent
asperity interlocking and collective stick-slip motion of the interface atoms, with a consequent
negligibly small frictional force.

Systems achieving low values of dry sliding friction are of great physical and potentially
technological interest, e.g., to significantly reduce dissipation and wear in mechanical devices
functioning at various scales. Superlubricity, with usually consequent ultra-low dynamic friction,
is experimentally rare, and has been demonstrated or implied in a relatively small number of cases,
such as sliding graphite flakes on a graphitic substrate \cite{dienwiebel,filippov,wijk}, cluster
nanomanipulation \cite{schwartz,schirmeisen}, sliding colloidal layers \cite{Bohlein12,VanossiNatMat,Vanossi12PNAS}, 
and inertially driven rare gas adsorbates \cite{mistura_nnano,cieplak,varini}.

Until very recently, superlubricity has been practically observed only on the nanoscale. A short time ago, a
breakthrough has been achieved, demonstrating the existence of superlubric regime of motion for
micrometer size graphite samples \cite{liu} and centimeter-long double walled carbon nanotubes (DWCNTs) \cite{zhang}. 
With the current capability of synthesizing and manipulating quasi 1D atomically perfect objects of extended length, such as telescopic
nanotubes \cite{zhang}, graphene nanoribbons \cite{fasel}, aromatic polymers \cite{kawai}, or soft biological filaments \cite{vitelli}, 
nanotechnologies open now the possibility to transpose the peculiar nanoscale tribological properties to larger scales and exploit
them to control sliding-induced energy dissipation in state-of-the-art technological devices.

Though, the robustness of the superlubricity phenomenon remains a challenge, and the conditions of
its persistence or the mechanisms of its failure are cast as key questions to be addressed. Even in
clean wearless friction experiments with perfect atomic structures, superlubricity at large scales
may surrender due to elasticity of contacting samples \cite{muser2004}.

A recent letter \cite{ming}, via a FK modeling approach, has shown that the effective rigidity of an
incommensurate slider, necessary condition for the occurrence of a superlubric regime, is strongly
affected by the driving protocol used to induce motion. In an edge-driven configuration, where the
external pulling or pushing force is applied to the edge of the slider, and which is typical for
many tribological and nanomanipulation experiments performed with atomic force microscope, a slider
critical length (depending on intrinsic material properties and experimental parameters) emerges
above which superlubricity breaks down. 
Here, we highlight new features of the transition from superlubric to high-friction state 
which occurs above the critical length. In particular, we consider different driving and commensurability configurations, 
discovering the occurrence of previously undetected multiple jumps in the friction force as a function of the slider length.
We analyze in details the mechanisms leading to the striking boost in dissipated energy materializing abruptly, above the critical size,
due to a localized deformation of the slider becoming commensurate to the underneath substrate
lattice and thus giving rise to dynamical high-friction stick-slip regimes.

The paper is organized as follows. In the next section, we present the main features 
of the edge-driven FK-type model, briefly summarizing the simulation procedure.
Sections III and IV are devoted to the numerical results elucidating the novel mechanisms 
of energy dissipation leading to the breakdown of superlubricity for long extended systems,
together with a consistent analytical framework for the determination of the length critical 
size. Sections V discusses relevant characteristics of the driving-induced commensurate 
dislocations in the slider, specifically in relation to their spatial localization and width depending on 
the system parameters. The novel occurrence of multiple commensurate walls is 
also presented. Conclusions are given in the last section.

\section{The model}
\begin{figure}
\includegraphics[width=8.5cm]{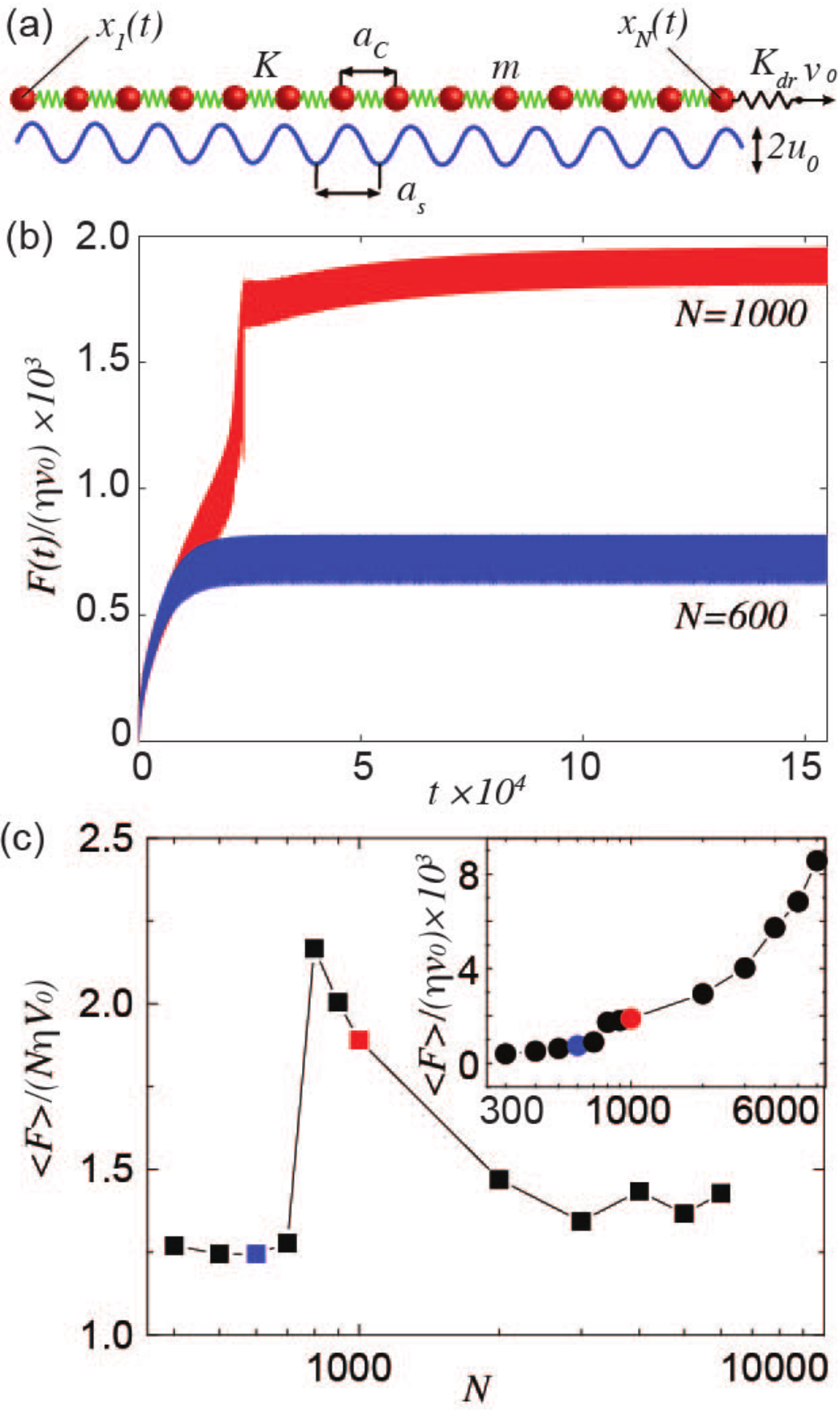}
\caption{(a) Sketch of the Frenkel-Kontorova-like model discussed in the work. (b) Instantaneous friction force normalized by the average single-particle viscous force for different chain lengths. (c) Average friction force per particle normalized by the viscous force as a function of the chain length, the colored squares refer to panel (b). The inset shows the average friction force as a function of the chain length. The parameter values used in these simulations are $u_0 = 0.02$, $\eta = 3.2\times 10^{-2}$, $v_0 = 1.6\times 10^{-2}$ and $k_{dr}= 0.1$.}
\label{figure1}
\end{figure}
The mechanical and tribological properties of the quasi 1D nanostructures previously introduced can be analyzed in the framework of the Frenkel-Kotorova model \cite{Braunbook,vanossiJPCM}. Specifically, they are treated as finite chains of $N$ particles of mass $m$, connected by springs of stiffness $K$, having rest length $a_c$, and driven on a sinusoidal potential with periodicity $a_s$ and amplitude $U_0$ representing the interaction with the substrate.
As sketched in Fig.\ref{figure1} (a) the rightmost particle of the chain, with coordinate $X_N(t)$, is pulled or pushed at constant velocity $V_0$ through a spring of constant $K_{dr}$ representing the lateral stiffness of the AFM cantilever. 
As in real AFM experiments the friction force $F(t)$ is estimated by the lateral deflection of the cantilever, i.e. by the spring elongation $F=k_{dr}[V_0 t - X_N(t)]$. A useful dimensionless unit system is obtained setting $a_c$ as the unit length, $\tau=\sqrt{m/K}$ as the unit time and $K a_c^2$ as the unit energy, with this choice the equations of motion are:
\begin{equation}
\begin{cases}
\ddot{x}_i=F_i^{sub}+F_i^{el}-\gamma \dot{x}_i   & for \; i<N\\
\ddot{x}_N= F_N^{sub}+F_N^{el}-\gamma \dot{x}_N+k_{dr}(v_0 t - x_N)& for \; i=N
\end{cases}
\label{newton}
\end{equation}
with the substrate force given by
\begin{equation}
F_i^{sub}=-\frac{2 \pi u_0}{a_s / a_c}\cos \bigg( \frac{2 \pi x_i}{a_s / a_c} \bigg),
\end{equation}
and elastic forces by
\begin{equation}
F_i^{el}=
\begin{cases}
x_{2}+x_{1}-1& for \; i=1\\
x_{i+1}+x_{i-1}-2x_i& for \; 1<i<N\\
x_{N}+x_{N-1}-1& for \; i=N
\end{cases}
\end{equation}
In the dimensionless unit system $x_i=X_i/a_c$, $u_0=U_0/(K a_c^2) $, $v_0=V_0(a_c/\tau)$ and $k_{dr}=K_{dr}/K$. The work done by the external force excites the internal degrees of freedom of the driven nanostructure first, e.g. molecular modes, and is subsequently dissipated through the substrate degrees of freedom, i.e. phonons. To dispose off the extra energy injected by the external driving a viscous damping term has been included in the equations,  
$\gamma=\eta \tau$ is the dimensionless viscous coefficient if $\eta$ is the dimensional one, $1/\eta$ represents the characteristic time necessary for the energy dissipation.

The equations of motion (\ref{newton}) have been integrated numerically using a velocity-Verlet algorithm. Finite temperature simulations have been also performed by means of a 
Langevin thermostat, however we found that up to $T=0.4 U_0/k_B$ the thermal effects do not lead to qualitative changes, thus we focus here on the case of T=0 only.

All the quasi 1D systems previously discussed present some common feature: (i) their periodic atomic structure is incommensurate with respect to the substrate one; (ii) their 
interatomic stiffness $K$ is sufficiently larger than the interfacial stiffness $K_{int}=U_0(2 \pi/a_s)^2$.  In the thermodynamic limit ($N \rightarrow \infty$), a 
continuum set of ground states that can be reached adiabatically through nonrigid displacement of chain atoms at no energy cost (Goldstone mode), with a consequent vanishing static 
friction \cite{aubry}.

\section{Breakdown of superlubricity}
\label{main}
\begin{figure*}
\includegraphics[width=1.\linewidth]{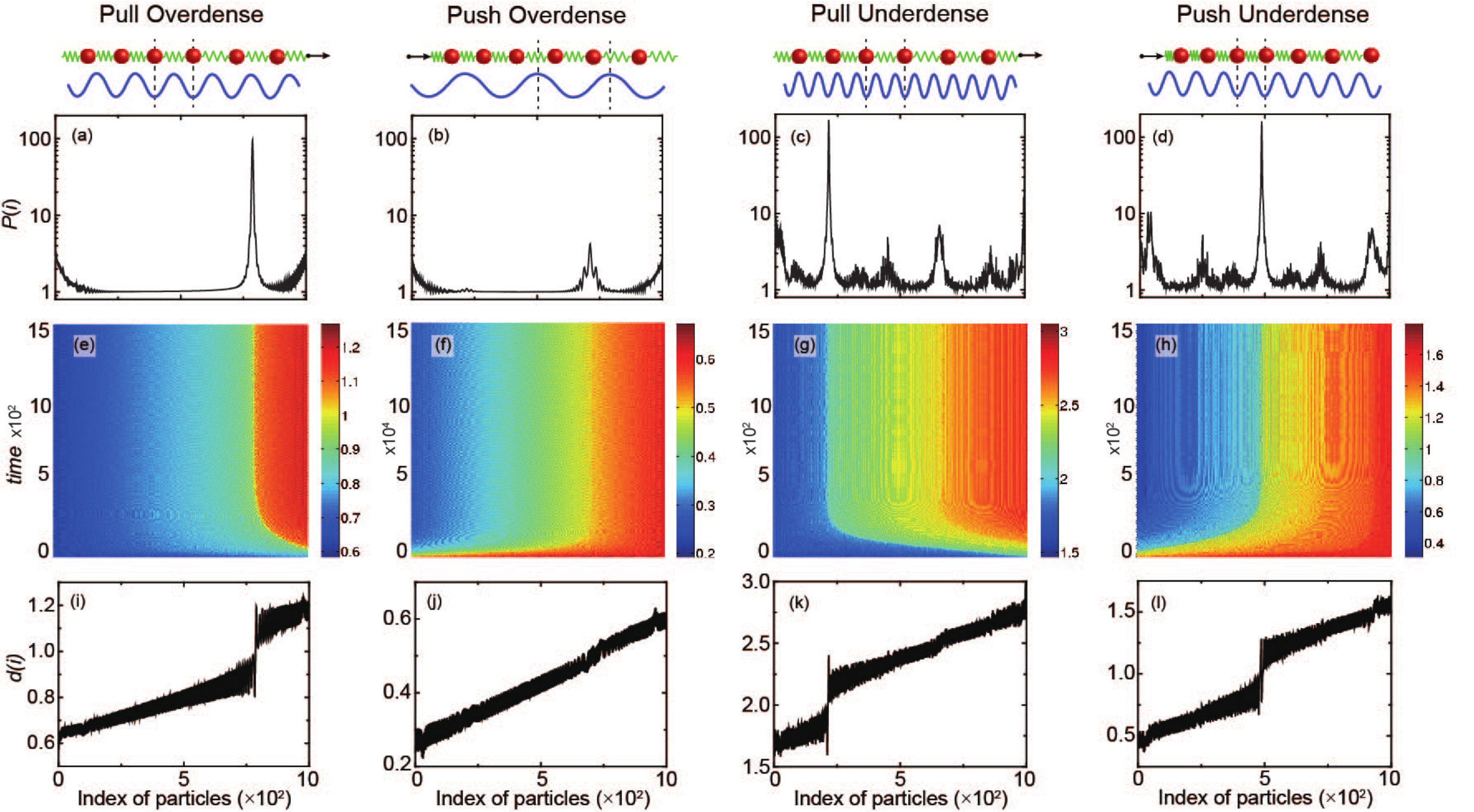}
\caption{Breakdown of superlubricity for a chain of $N=1000$ particles subjected to a pulling or pushing driving in overdense or underdense condition. (a)-(d) average power dissipated $P(i)$ by the $i$-th particle as a function of the particle position in the chain, the inset schematize the arrangement of the particles in the wall region. (e)-(h) color maps showing the behavior of the commensuration index $d(i)$ in time. (i)-(l) two dimensional plots of $d(i)$ in the steady sliding state. 
The model parameters are the same of fig.\ref{figure1}, for the underdense case we have chosen $a_c/a_s=(1+\sqrt{5})/2$ while for the overdense one $a_c /a_s=2/(1+\sqrt{5})$.}
\label{figure2}
\end{figure*}

Before discussing in detail the novel mechanism of dissipation leading to the breakdown of superlubricity for long chains, we start analyzing the frictional response as a function of the chain length first considering an overdense chain ($a_s/a_c=(1+\sqrt{5})/2$) under pulling driving. The blue curve of fig.\ref{figure1} (b) represents the typical friction force trend with an initial transient, due to the stretching of the chain, and the onset of a plateau where friction oscillates around constant average value. For very short chains edge effects might be predominant giving rise to a large average friction, only when $N>N_e$ these effects become negligible and a super-low friction state onsets, with our choice of parameters $N_e\simeq 200$. At a critical particle number $N_{cr}\simeq 750$ a sudden increase of the average friction force breaks the super-low friction regime. The time evolution of the friction force for a chain with $N=1000>N_{cr}$ is reported in fig.\ref{figure1} (b) showing an abrupt jump during the initial transient almost doubling the average friction value in the steady sliding state plateau. A quantity intimately connected to friction is the energy dissipation, to better understand which changes affect the sliding motion for $N>N_{cr}$ one can analyze how the power dissipated by the viscous damping is distributed along the chain. More precisely, in our dimensionless unit system, one can plot a normalized time-averaged power dissipated by the $i$-th atom:
\begin{equation}
P(i)=\frac{1}{V_0^2} \lim_{T\rightarrow\infty}\frac{1}{T}\int_{0}^{T}\dot{x}_i^2 dt
\end{equation}
where, in the time interval $T$ over which the average is taken, the chains already reached the steady sliding state. For $N<N_{cr}$ such a quantity is almost zero everywhere except at the edges, due to a slightly higher particle mobility but, as shown in fig.\ref{figure2} (a), for $N>N_{cr}$ a narrow region appears, in a specific point of the chain, where dissipation increases by two orders of magnitude. A different chain arrangement must evidently give rise locally to a completely different and extremely dissipative sliding mechanism. 
To further elucidate the local rearrangement of the particles it is convenient to plot the local commensuration index:  
\begin{equation}
d(i)=\frac{X_i-X_{i-1}}{a_s},
\end{equation}
i.e. a quantity close to one when two neighboring particles are locally in registry with the substrate and larger or smaller than one if the two particles are closer or far apart than $a_s$. 
A color map representing the time evolution of the commensuration index, from the initiation of driving to the onset of the steady sliding state, is plotted in fig.\ref{figure2} (e). 
A narrow yellow region where nearby particles are commensurate ($d(i)\simeq 1$) nucleates at the driving edge and propagates backward in the chain upon reaching a certain constant position $L_{cr}$. It works as a domain wall separating two incommensurate chain portions with $d(i)<1$ on the left and $d(i)>1$ on the right. Fig.\ref{figure2} (i) shows a 
cross section of the 2D color map taken at a large simulation time where the steady state is already reached, it clearly shows that the chain undergoes a linear stretching with a discrete jump in correspondence of the domain wall. It is easily understandable, and readily verified looking directly at the particle trajectories, that stick-slip motion insets within the wall region with a consequent enhancement of dissipation and loss of superlubricity.

Changing the driving mode from pulling to pushing, a jump in the friction force is also observed. As shown in fig.\ref{figure2} (f) this corresponds to the nucleation, at the trailing edge, of a narrow domain wall region in which $d(i)\simeq 0.5$, i.e. a region of local commensuration where the compressed particles assume a periodicity half of the substrate lattice spacing $a_s$. As in the pulling case this domain wall region, separating two incommensurate domains having  $d(i)<0.5$ on the left and $d(i)>0.5$, propagates forward in the chain up to a certain location $L_{cr}$. Fig.\ref{figure2} (b) shows the occurrence of a dissipation peak in analogy to the pulling case. The overall friction force experienced by the chain is smaller than in the pulling counterpart, while a $1/1$ commensuration requires in fact the wall particles to sit in the very bottom of the surface potential, with a consequent large effort to make them sliding forward, in a $0.5/1$ commensuration the two 
particles sharing the same potential valley will lie far from the minimum, less force is needed to push the wall forward.

Up to now we only considered overdense chains but, given the asymmetry between pushing and pulling driving, it is definitely interesting to investigate also the behavior of underdense chains, i.e. the case $a_c/a_s=(1+\sqrt{5})/2$.
Also in this cases pushing and pulling give rise to very different behaviors as shown by fig.\ref{figure2} (g) and (h). Again we have the nucleation and propagation of domain walls where the particles get commensurate with respect to the substrate potential, however, given the lower density of particles in the chain, upon pulling we reach a $2/1$ commensuration while upon pushing the particles are squeezed back to a $1/1$ configuration, see fig.\ref{figure2} (k) and (l). In both cases the commensurate particles live in the very bottom of the surface potential and, as in the overdense pulling case, the total friction force jumps by more than a factor two for $N>N_{cr}$. Looking at the average dissipated power, shown in fig.\ref{figure2} (c) and (d), one might notice again a large peak in correspondence of the domain wall region, but also a pronounced dissipation elsewhere, contrary to the overdense analogs. In the underdense case this can be explained by a much larger variety of partially commensurate slider-substrate configurations, leading to a broader multiplicity of dissipation peaks.   

\section{Analytic treatment}
\begin{figure}
\includegraphics[width=1.\linewidth]{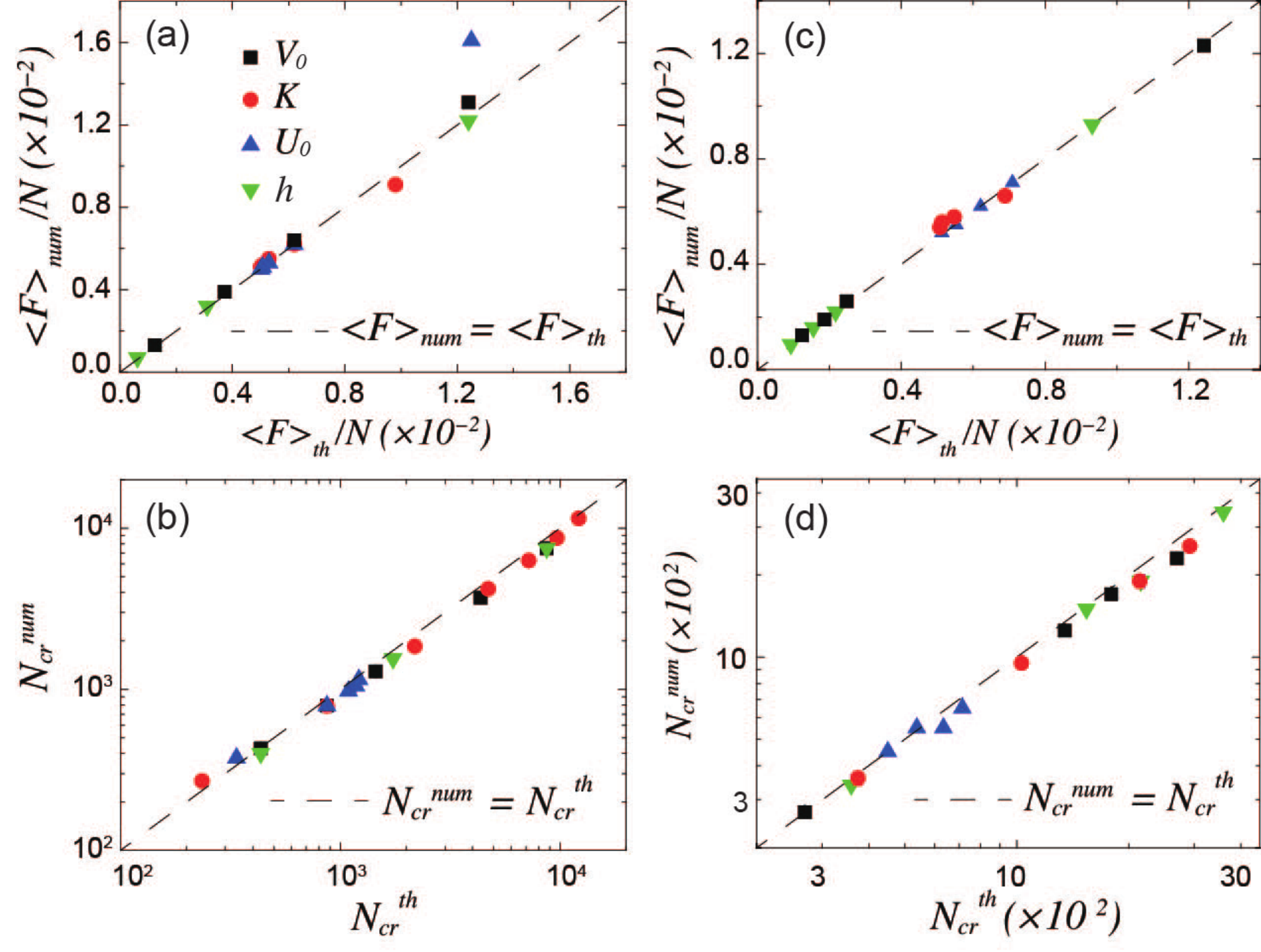}
\caption{
(a),(c) Comparison between the average friction force per particle $\langle F \rangle/N$ calculated numerically and given by eq.(\ref{fit}) for the pulling driving in the 
overdense and underdense case respectively. The fitting parameter is $\alpha_1=2.64\pm 0.11$ for the overdense case and $\alpha_1=2.69\pm 0.15$ for the underdense
one. (b),(d) Comparison between the critical number of particles $N_{cr}$ calculated numerically and given by eq.(\ref{eccola}) for the pulling driving in the overdense and 
underdense case respectively.
Black squares, red circles, blue and green triangle are obtained by varying one parameter at a time in the range $0.02<U_0<0.5$, $5<K<100$, $0.005<V_0<0.1$ and $0.01<\eta<0.2$ 
keeping the others set to $U_0=0.2$, $K=10$, $V_0=0.05$ and $\eta=0.1$. The dashed black lines are the bisectors of the plots, i.e. the set of points where $\langle F \rangle_{num}=\langle F \rangle_{th}$ and $N_{cr}^{num}=N_{cr}
^{th}$.}
\label{figure5}
\end{figure}
In this section we derive an analytic expression for the critical number of particles in the chain $N_{cr}$ above which superlubricity is broken. The explicit calculation
will be carried out for the pulling of overdense chains only, the same arguments can be straightforwardly applied to the other three cases above presented.\\
We start writing the time-average friction force for a generic chain in its steady sliding state \cite{elmer}:
\begin{equation}
\langle F\rangle=m\eta N v_0 \bigg[ 1 + \lim_{T\rightarrow \infty}\frac{1}{T} \int_{t_0}^{T} \frac{1}{N}\sum_{j=1}^{N}\bigg(  \frac{\dot{x}_j}{v_0}-1\bigg)^2 dt \bigg].
\label{eb}
\end{equation}
This equation is exact and is derived simply imposing energy conservation, i.e. the energy pumped into the chain by the external driving per unit time equals the energy dissipated per unit time by the viscous damping applied to the particles
\begin{equation}
m\eta  \lim_{T\rightarrow \infty}\frac{1}{T} \int_{t_0}^{T} \frac{1}{N}\sum_{j=1}^{N}\dot{x}_j^2 dt.
\end{equation}
For chains with $N<N_{cr}$, the absence of stick-slip allows us to apply a linear perturbation theory writing $\dot{x}_i=\dot{x}_i^0+\epsilon \dot{x}_i^1+O(\epsilon^2)$, here $\epsilon=K_{int}/K$ is a small dimensionless parameters
representing the ratio between the chain and substrate stiffness, $\dot{x}_i^0=v_0$ is the exact solution in absence of the substrate potential, i.e. $\epsilon=0$. Substituting the velocity expansion in eq.(\ref{eb}) we get \cite{elmer}
\begin{equation}
\langle F\rangle=m\eta N v_0 \bigg[ 1 + \alpha_1 \bigg( \frac{K_{int}}{K}\bigg)^2+ O(\epsilon^3)\bigg],
\label{fit}
\end{equation}
all the unknowns hidden inside $\alpha_1$, while it remains hard to get a simple analytic expression for this parameter, we can easily find its numerical value by fitting the simulation 
results. To this aim we performed many different calculations of the average friction force varying $U_0$, $K$, $V_0$ and $\eta$ for chains with number of particles $N_b<N<N_{cr}$ 
where $\langle F\rangle/N$ is almost independent of $N$, i.e. in the initial plateau of fig.\ref{figure1} (c). Choosing $\alpha_1=2.64\pm 0.11$ we obtain a remarkable agreement 
between the analytic expression (\ref{fit}) and the numerical results with $0.03 <\epsilon <0.3$ as shown in figure \ref{figure5} (a). Only for large $U_0$, namely $\epsilon=0.75$, 
we start to see a significant discrepancy between the model and the simulations, for such a large $\epsilon$ value the adopted linear expansion is clearly insufficient and higher order 
therms should be included in eq.(\ref{fit}).\\
Once again we stress that eq.(\ref{fit}) describes a viscous-like friction, proportional to $V_0$ and $\eta$ and only weakly dependent on $U_0$ and $K$ and is suitable for $N<N_{cr}$. 
For $N>N_{cr}$ the domain wall region nucleating inside the driven chain is characterized by a stick-slip type of friction, increasing with $U_0$ and $K$ and practically independent of $V_0$ and $\eta$, such an highly non-linear kind of motion cannot be approximated with a power expansion. Our simulations show that above the critical length the friction force can be calculated as a sum of the contributions given by eq.(\ref{fit}) and by the dissipation in the local region at the domain wall.\\
\begin{figure}
\includegraphics[width=1.\linewidth]{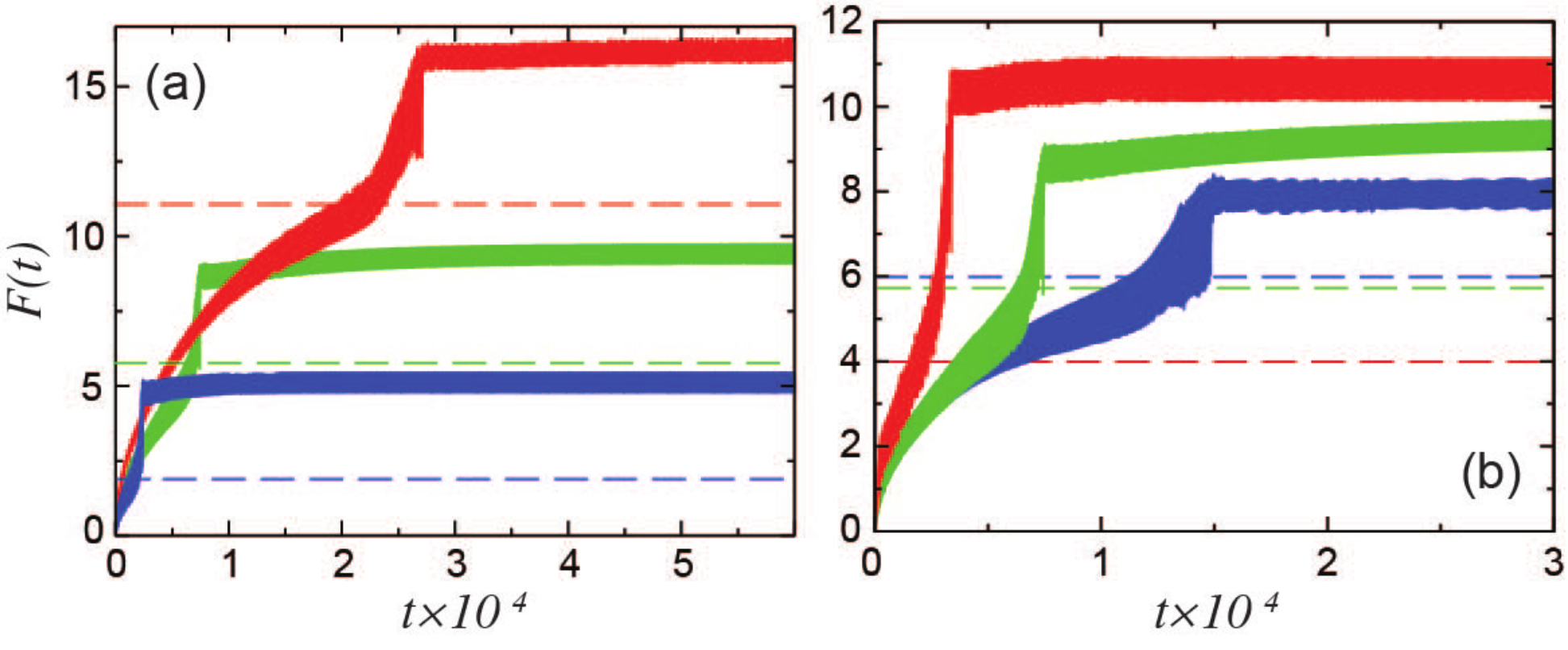}
\caption{Comparison between the simulated $F(t)$ and the theoretical prediction of $F_{ext}^{cr}$ from eq.(\ref{crit}) for different values of (a) $K$, (b) $U_0$. The curves show numerical results for $F(t)$ and the dashed lines are the theoretical values from eq.(\ref{crit}). (a) The red, green and blue curves are calculated for $K= 20$ $N = 2000$, $K= 10$ $N = 2000$ and $K = 5$ $N = 360$ respectively. (b) The red, green and blue curves correspond to $U_0 = 0.5$ $N = 450$, $U_0 = 0.2$ $N = 1000$ and $U_0 = 0.1$ $N = 1000$ respectively.}
\label{figure3}
\end{figure}
As discussed in the previous section the maximum length at which the driven chains exhibit a super-low friction behavior is limited by the nucleation of a localized commensurate region at the driving edge. The external force $F_{ext}$ necessary to commensurate the $N$ and $N-1$ particles with the underneath substrate potential can be easily calculated
\begin{equation}
F_{ext}=K(a_s-a_c)+\frac{2 \pi u_0}{a_s}\cos \bigg( \frac{2 \pi x_N}{a_s}\bigg),
\end{equation}
simply being the sum of the force necessary to stretch the first spring from its rest length $a_c$ to the substrate periodicity $a_s$ and the force payed by moving the $N$-th particle on the substrate potential. This expression depends of course on the absolute position of the chain edge but its smallest value, i.e. the smallest value $F_{ext}^{cr}$ necessary for the nucleation of the wall is:
\begin{equation}
F_{ext}^{cr}=K(a_s-a_c)-\frac{2 \pi u_0}{a_s},
\label{crit}
\end{equation}
obtained when the $N$-th particle sits on a maximum. Clearly for the other three cases presented in fig.\ref{figure2} the nucleation condition, to be imposed in order to calculate $F_{ext}^{cr}$, is different as the domain wall region is constituted by a different arrangement of the particles, as schematized by the cartoon insets of panels (a)-(d).  
Expression (\ref{crit}) is exact and can be directly verified through the simulations. As shown in Fig.\ref{figure3} the $F_{ext}^{cr}$ value calculated from eq.(\ref{crit}) for different $K$ and $U_0$, dashed lines, always match the value at which $F(t)$ jumps vertically in the corresponding simulations, i.e. the force value at which the commensurate region nucleates.
Moreover, according to eq.(\ref{crit}), the critical force does not depend on the length of the chain and on the dynamical parameters, i.e. the viscous damping and the driving velocity. This is also easily verified through our simulations, Fig.\ref{figure4} shows again $F(t)$ during the nucleation process for different $\eta$, $V_0$ and $N$ and is clearly seen that, although the nucleation takes place at different times, $F_{ext}^{cr}$ is constant.\\ 
Now, in a steady sliding regime the time-averaged pulling force equals the average friction force and an expression for the critical number of particle $N_{cr}$ can thus be obtained equating (\ref{fit}) and (\ref{crit})
\begin{equation}
N_{cr}=\frac{K(a_s-a_c)}{m\eta v_0}\bigg( 1-\beta \frac{K_{int}}{K}\bigg)\bigg[ 1 + \alpha_1 \bigg( \frac{K_{int}}{K}\bigg)^2\bigg]
\label{eccola}
\end{equation}
with $\beta=a_s/[2 \pi (a_s-a_c)]$. This new expression shows that the critical number of particles increases with the
stiffness of the chain and decreases with increasing the damping coefficient and pulling velocity. $N_{cr}$ also decreases with increasing the ratio between the two stiffness, $K_{int}/K$, 
however this effect is weak since $K_{int}/K$ is a small parameter. To verify the robustness of our model we tested the prediction of eq.(\ref{eccola}) against the numerical simulations 
performed with a broad range of parameters $V_0$, $\eta$, $U_0$ and $K$, the results are shown in fig.\ref{figure5} (b). For all values of system parameters the theoretical results 
agree well with numerical simulations. This result show that the critical number of particles is mostly determined by the stiffness of the chain $K$, the damping coefficient $\eta$, and pulling 
velocity $V_0$ rather than by the ratio $K_{int}/K$ as suggested in previous theoretical works \cite{muser2004}.

In this work, to magnify and highlight the breaking of superlubricity, we only considered the incommensurability of the chain and substrate structures corresponding to the golden ratio, 
for which the difference between two periods is relatively large. In real systems with such a large period mismatch anharmonic effects are expected to play a role, but we believe they do 
not alter the overall picture obtained within the linear elasticity approximation adopted in our numerical and analytic models. Moreover, the results here discussed 
are valid also for smaller misfits between the contacting lattices. Then the stretching of the chain at the transition is significantly smaller and anharmonic effects can be neglected again. 
\begin{figure}
\includegraphics[width=1.\linewidth]{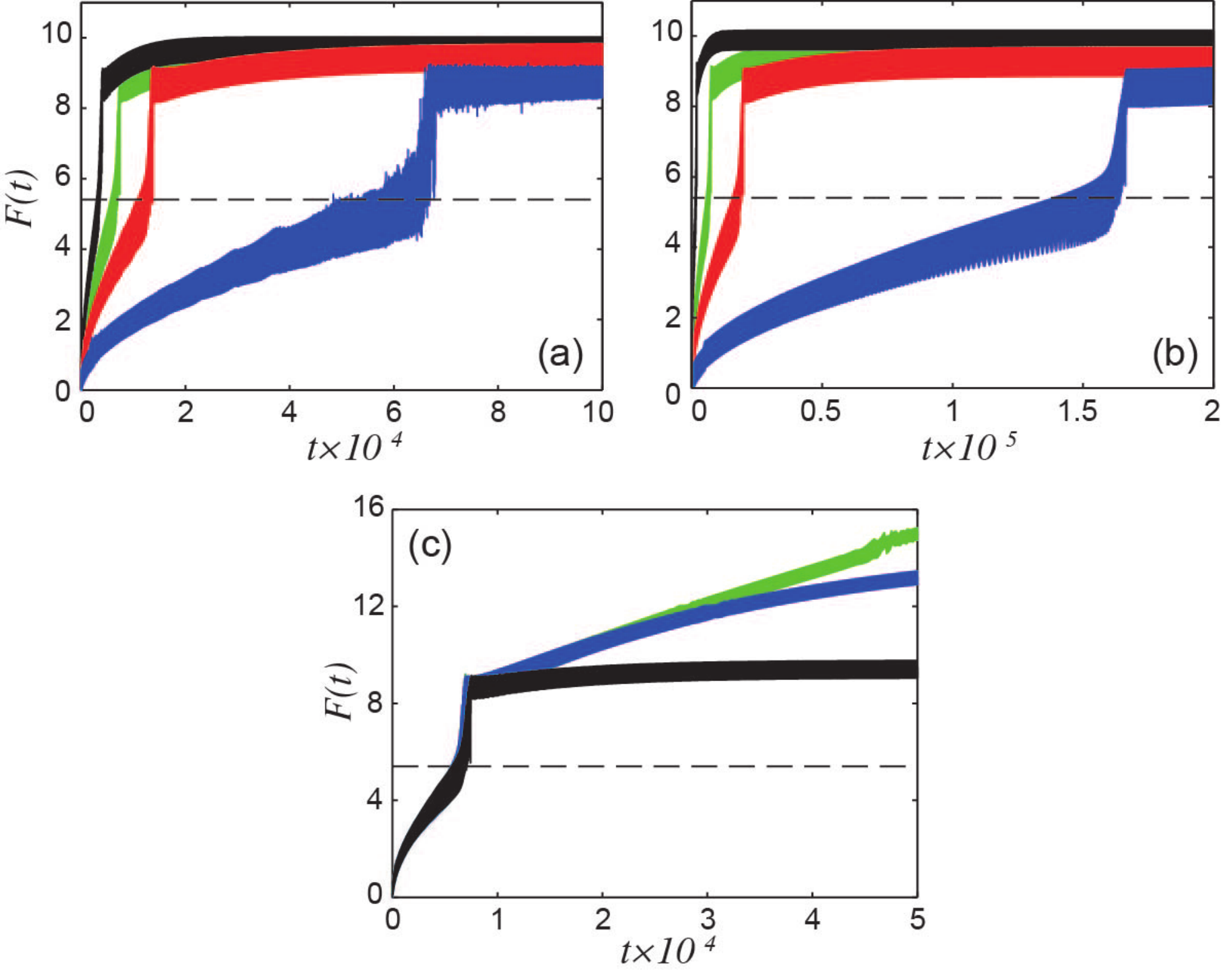}
\caption{Comparison between the simulated $F(t)$ and the theoretical prediction of $F_{ext}^{cr}$ from eq.(\ref{crit}) for different values of (a) $\eta$, (b) $V_0$ and (c) $N$. The curves show numerical results for $F(t)$ and the 
dashed lines are the theoretical values from eq.(\ref{crit}). (a) the black, green, red and blue curves are calculated for $\eta = 0.2$ $N = 540$, $\eta = 0.1$ $N = 1000$, $\eta = 0.05$ $N = 2000$ and $\eta = 0.01$ $N = 9000$ respectively. (b) the black, green, red and blue curves correspond to $V_0 = 0.1$ $N = 560$, $V_0 = 0.05$ $N = 1000$, $V_0 = 0.03$ $N = 1600$ and $V_0 = 0.01$ $N = 4600$ respectively. (c) the black, blue and green curves correspond to $N = 1000$, $2000$ and $100000$ respectively.}
\label{figure4}
\end{figure}
Finally it should be noted that our calculations of
$N_{cr}$ are based on the assumption that all particles of the chain experience the same microscopic friction proportional to the viscous damping coefficient. However, in 
realistic systems there could be additional contributions to friction coming from chemical interactions between the edge particles and the surface and/or defects in the chain structure. 
These effects will lead to an increase of the average friction force thus reducing the critical number of particles, which could benefit the experimental investigations of the mechanisms discussed here.\\
Repeating the same fitting procedure for the underdense condition of $a_c/a_s=(1+\sqrt{5})/2$, we find $\alpha_1=2.69\pm 0.15$ the same value as before 
within the estimated error. Fig.\ref{figure5} (c) and (d) report the comparison between the analytic model prediction and the numerical simulations for the average friction force and $N_{cr}$.
 
\section{Domain wall features}
\begin{figure}
\includegraphics[width=1.\linewidth]{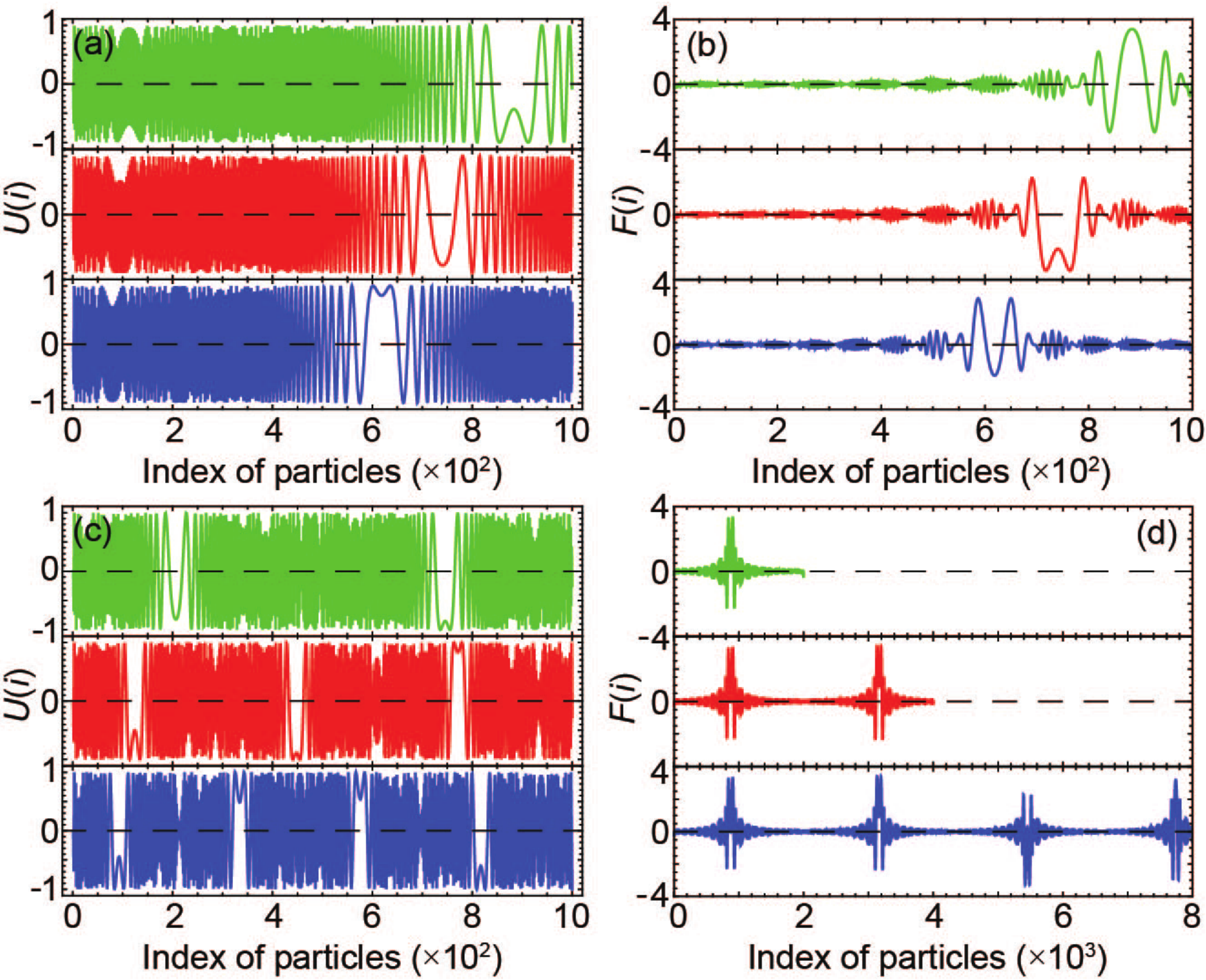}
\caption{(a) Single particle potential energy $U(i)$ as a function of the particle index for a pulled overdense chain with $N=1000$. The three different plots correspond to $
\delta=7.0\times10^{-4}$ (green), $\delta=8.35\times10^{-4}$ (red) and $\delta=1.0\times10^{-3}$ (blue). (b) Single particle force $F^{sub}(i)$ obtained differentiating the 
curves of panel (a), to cancel the incoherent contributions and highlight only the regions where the force is significantly different from zero, the force value at every point $i$ as been 
obtained averaging over the five neighboring points (average filtering). (c) Single particle potential energy $U(i)$ for larger $\delta$ values showing the onset of multiple domain walls with a periodicity inversely proportional to $\epsilon$.  
The three different plots correspond to $\delta=3.0\times10^{-3}$ (green), $\delta=5.0\times10^{-3}$ (red) and $\delta=7.0\times10^{-3}$ (blue). (d) single particle force $F^{sub}(i)$ for $\delta=7.0\times10^{-4}$ and different chain lengths. The three different plots correspond to $
N=1000$ (green), $N=4000$ (red) and $N=8000$ (blue).}
\label{figure6}
\end{figure}

\begin{figure}
\includegraphics[width=0.5\linewidth]{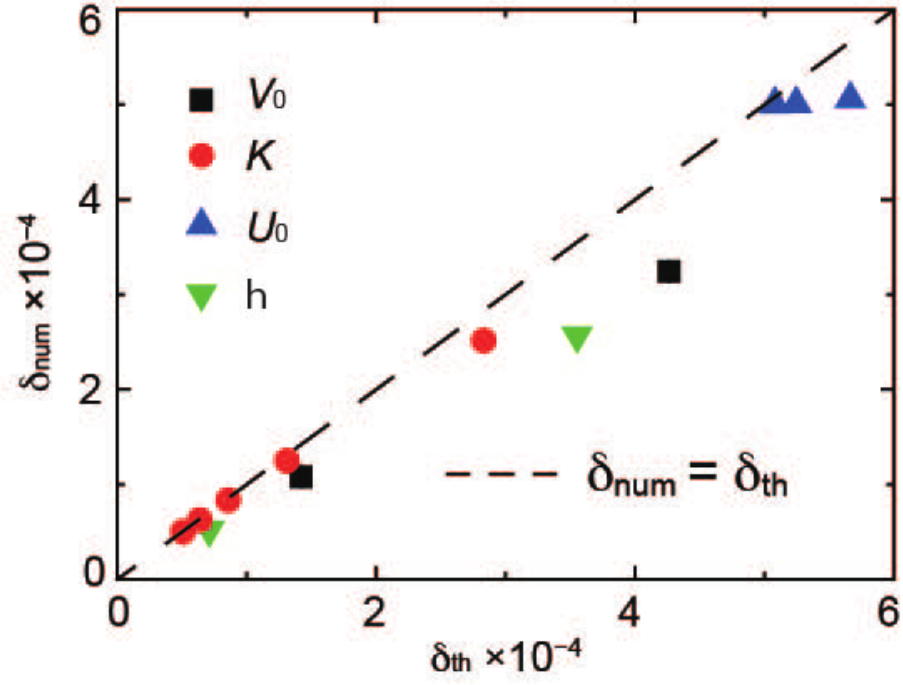}
\caption{Comparison of the numerically calculated slope $\delta$ with the theoretical prediction of eq.(\ref{sega}) with $N_{cr}$ corresponding to the theoretical value of eq.(\ref{eccola}). Black squares, red circles, blue and green triangles obtained by varying $0.01<V_0<0.03$, $20<K<100$, $0.02<U_0<0.1$ and $0.01<\eta<0.05$ respectively while keeping the other parameters constant at the values $U_0=0.2$, $K=10$, $V_0=0.05$ and $\eta=0.1$.}
\label{figure7}
\end{figure}

\begin{figure}
\includegraphics[width=1.\linewidth]{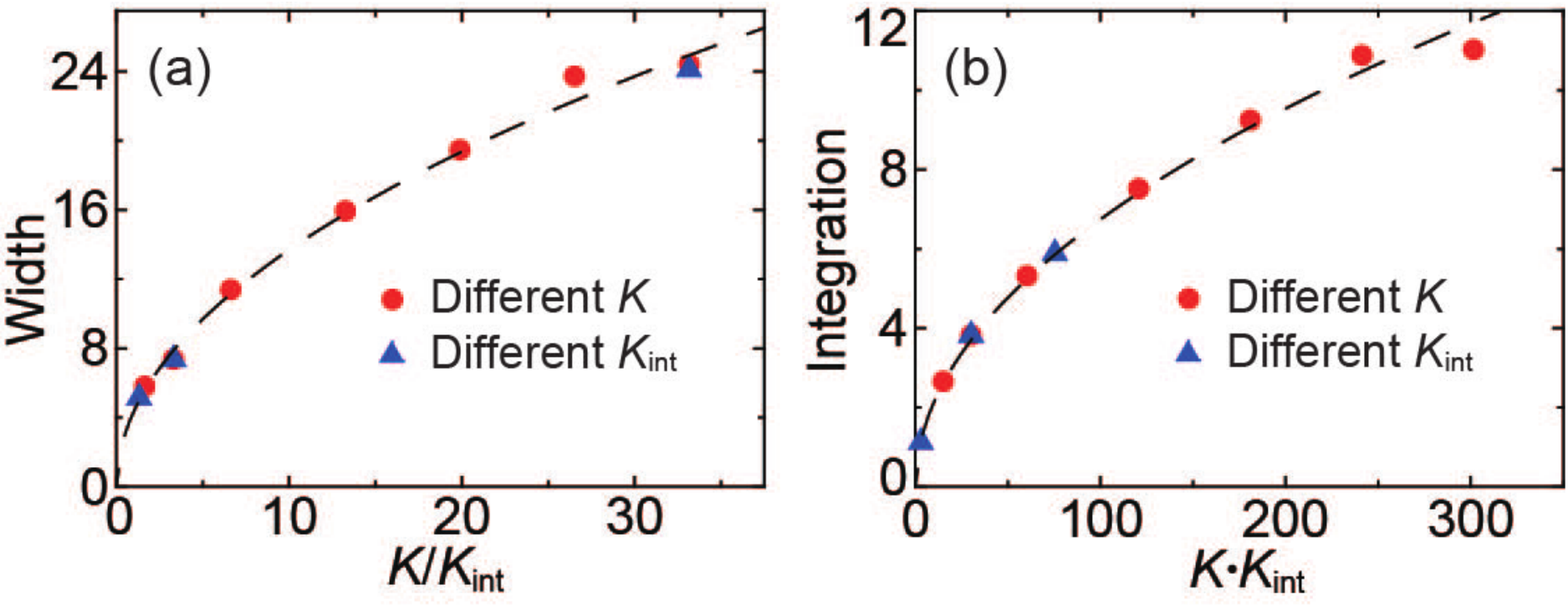}
\caption{(a) Width of the dissipation peak of fig.\ref{figure2} (a) as a function of the chain and substrate stiffness. (b) Integral dissipated power as a function of the chain and substrate stiffness. Red circles and blue triangles are obtained by varying $5<K<100$ and $0.02<U_0<0.5$ respectively while keeping constant the other parameters at the values $U_0=0.2$, $K=10$, $V_0=0.05$ and $\eta=0.1$.}
\label{figure8}
\end{figure}

\begin{figure}
\includegraphics[width=1.\linewidth]{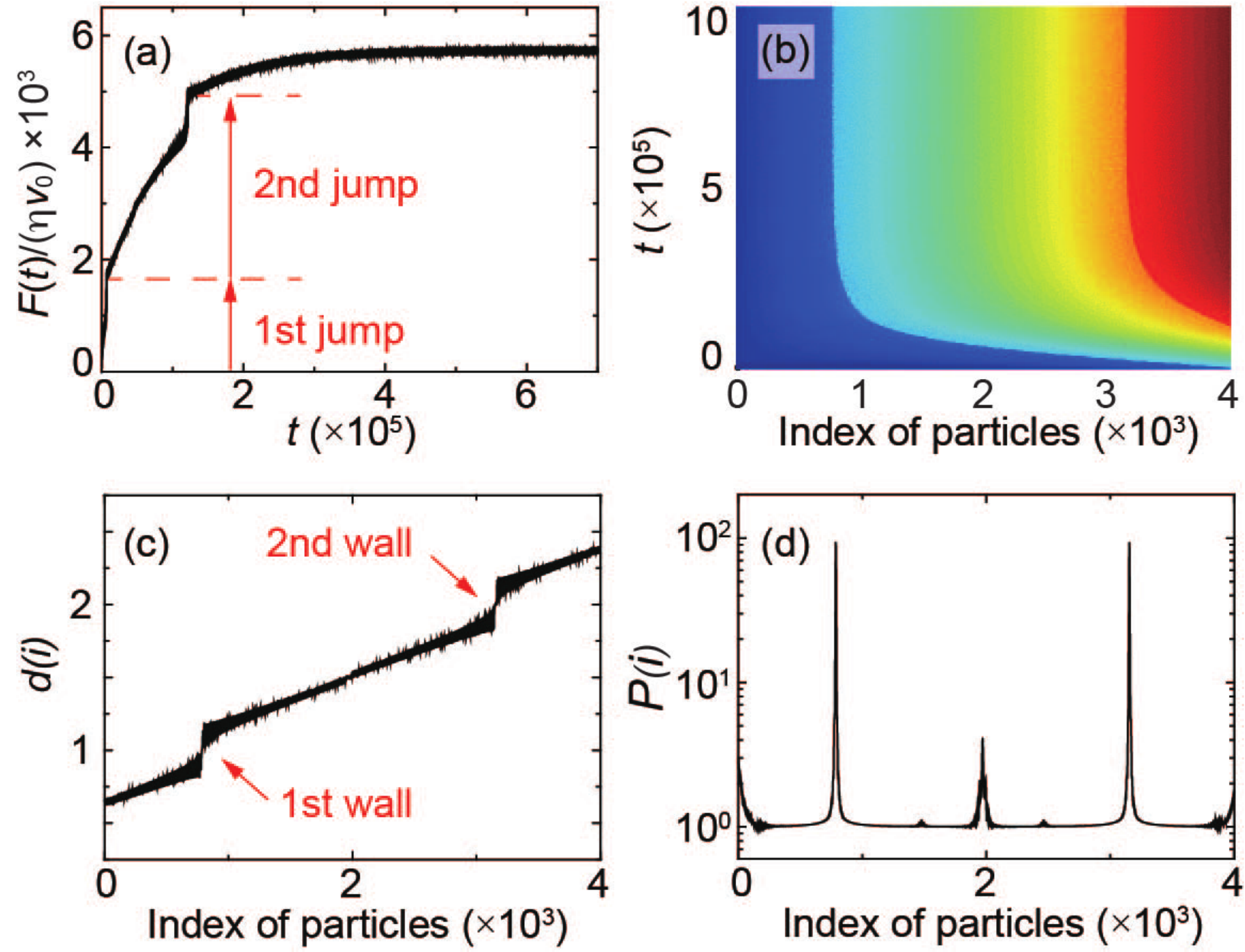}
\caption{Onset of a second domain wall during the pulling of a long overdense chain. (a) jumps in the total friction force as the commensurate domain wall regions nucleates at 
the driving edge. (b) color map showing the time evolution of the commensuration index $d(i)$. (c) commensuration index $d(i)$ in the steady sliding state. (d) average power 
dissipated $P(i)$ by the $i$-th particle as a function of the particle position in the chain. The parameters used in this simulation are the same as in fig.\ref{figure1} and \ref{figure2}, only 
the total length has been increased up to $N=4000$.}
\label{figure9}
\end{figure}

As discussed in section \ref{main} and immediately visible in fig.\ref{figure2}, upon reaching the steady sliding state, the narrow domain wall region takes a well defined and constant 
position $L_{cr}$ along the chain. This position is determined by the chain and substrate stiffness and by the dynamic parameters $m$, $\eta$,$v_0$. 
On the contrary $L_{cr}$ is not affected by an increase of the chain length, upon increasing $N$ the 
distance between the free edge and the domain wall is unchanged while the distance between the domain wall and the driven edge increase linearly. The only other quantity 
displaying such a behavior is the chain stretching, i.e. the slope of the curves in fig.\ref{figure2} (i)-(l), influenced by substrate and chain stiffness and by the driving conditions but 
independent of $N$ both above and below $N_{cr}$. 
It is thus clear that, once the driving dependent stretched state is reached, $L_{cr}$ is solely determined by the 
interplay between the periodic substrate potential and the linearly increasing periodicity of the chain. The position of the $i$-th particle in a linearly stretched chain is given 
by $X_i = a_c(i - 1) +\delta (i - 1)i/2$, where the slope of linear stretching $\delta=\delta(k,v_0,\eta,U_0,m)$ sets the overall length of the stressed 
chain, once the steady sliding state in reached, as a function of the driving parameters. The potential and force felt by every particle due to the interaction with the substrate potential is 
thus readily obtained:
\begin{eqnarray}
\label{periodo}
&U(i)=U_o \sin\bigg[2\pi \bigg(\frac{a_c(i - 1)}{a_s} + \delta \frac{i (i - 1)}{2 a_s}\bigg)\bigg],\\
&F^{sub}(i)=- \frac{2\pi U_o}{a_s} \cos\bigg[2\pi \bigg(\frac{a_c(i - 1)}{a_s} + \delta \frac{i (i - 1)}{2 a_s}\bigg)\bigg].
\end{eqnarray}
$U(i)$ is plotted in fig.\ref{figure6} (a) for increasing $\delta$ values in the overdense puling case: upon increasing the stretching of the chain a plateau region appears, i.e. a region 
where nearby particles feel the same substrate potential meaning that they are almost perfectly commensurate. Keeping increasing $\delta$ the commensurate region moves backward, 
in our simulations this backward motion stops when the steady sliding state is reached and no further stretching/elongation of the chain occurs. 
Elsewhere in the chain the potential felt is rapidly and incoherently oscillating, upon averaging over few neighboring particles the resulting force will thus be zero everywhere except around 
the commensurate region as show in fig.\ref{figure6} (b). An expression for $N_{cr}$ can thus be obtained as a function of $\delta$ for the pulled overdense chain imposing $X_{N}-X_{N-1}=a_s
$, this yields:
\begin{equation}
N_{cr}=1 + \frac{a_s - a_c}{\delta},
\label{sega}
\end{equation}
substituting into the $x_i$ equation we get the distance of the domain wall from the trailing edge:
\begin{equation}
L_{cr}=\frac{(a_s - a_c) (a_c + a_s + \delta)}{2 \delta}.
\label{star}
\end{equation}
From the simulated chain length in the stretched steady sliding state we can estimate $\delta$ and use it the two previous equations to evaluate $N_{cr}$ or $L_{cr}$. 
The red curve in fig.\ref{figure6} (a) has been obtained with 
$\delta=8.35\times10^{-4}$ corresponding to the chain of fig.\ref{figure2} (a), indeed the position of the plateau in the potential 
coincides with the position of the domain wall, the estimated critical number of particles $N_{cr}=741$ is in good agreement with the simulations, see fig.\ref{figure1} (c). 
One might be tempted to compare eqs. (\ref{sega}) and (\ref{eccola}) to obtain an analytical 
expression for $\delta(k,v_0,\eta,U_0,m)$. Fig.\ref{figure7} shows the comparison between such a theoretical prediction and the numerical values coming from
the simulations. The agreement is fairly good considering that eq.(\ref{sega}) is exact while eq.(\ref{eccola}) comes from a power expansion for small $U_0$ only.\\
As previously discussed and shown in fig.\ref{figure2}, the dissipated power $P(i)$ exhibits a high peak localized in the commensurate region of the chain, where it takes a value two orders of magnitude higher than in the rest of the chain. The appearance of this peak is the finger print of transition from the super-low to high friction when $N>N_{cr}$. 
Fig.\ref{figure8} (a) shows that, in the range of parameters here considered, the width of the dissipation peak scales as $\sqrt{K/K_{int}}$. The particles located in the narrow commensurate region 
corresponding to the domain wall perform stick-slip motion, and their dissipated energy is proportional to the amplitude of the particle-substrate interaction $U_0$. As 
a result, the integral power dissipated by the particles located in the commensurate region scales as $\sqrt{K \;K_{int}}$. 
Fig.\ref{figure8} (b) shows that this prediction is again in good agreement with the numerical simulations from our simulations, notice that this integral 
dissipated power is independent of the viscous damping coefficient $\eta$ and driving velocity $V_0$. The power dissipated by the particles located 
outside the domain wall is proportional to $\eta V_0$ and only slightly dependent on the parameters $K$ and $K_{int}$, see eq.(\ref{fit}). 
Thus, the mechanisms of energy dissipation in the commensurate and incommensurate domains of the chain are completely different, originating 
contributions to the net friction force that can differ by several orders of magnitude.\\
If, as just demonstrated, the occurrence of a commensuration region in the chain is due to a local matching between the substrate periodicity and 
the linearly increasing chain periodicity, it becomes essential to check whether this commensuration occurs only once or periodically along the chain. Fig.\ref{figure6} (c) shows the potential energy per 
particle as $\delta$ is increased further after the occurrence of the first plateau, indeed many other plateaus appear and condensate within the chain. One can also fix $\delta$ 
increasing the chain length as in fig.\ref{figure6} (d) (the average force is plotted in place of the potential in order to better appreciate small variations on a large $N$ interval)
force peaks (i.e. plateaus in the potential) appear regularly as $N$ grows in a sort of Moire pattern with $\delta$ dependent periodicity. 
As a consequence, for chains long enough and under the proper driving conditions, more domain wall regions are expected to nucleate leading to higher discrete jumps in the friction 
force. This behavior is illustrated in fig.\ref{figure9} obtained simulating a chain under the same driving conditions and parameters of fig.\ref{figure2} (a) but four times longer. Two 
identical domain wall regions propagate in the chain causing a considerable enhancement of friction and giving rise to localized dissipation peaks due to stick-slip sliding. 
Notice that the second jump occurs when the value of the pulling forces reaches the critical value
\begin{equation}
F_{ext}^{cr}=K(2 a_s-a_c)-\frac{2 \pi u_0}{a_s}.
\end{equation}
The corresponding average force exerted by the substrate potential is given by the red plot in fig.\ref{figure6} (d), also in this case the position of the average force peaks (read potential plateaus) corresponds nicely with the domain wall location along the chain.\\
A similar treatment could be done for the other three cases discussed in fig.\ref{figure2} of pushed and underdense chains, naturally the condition to be imposed to derive eqs. (\ref{star}) and (\ref{sega}) varies from case to case.
 
\section{Critical length estimation for experimental systems}
Before concluding we give a practical example of how eq.(\ref{eccola}) can be used to estimate the critical length for different systems whose tribological properties are experimentally accessible.
Among other, particularly interesting is the case of DWCNTs where super-low friction has been observed up to the centimeter length-scale \cite{zhang}.
For these systems the intrashell and intershell stiffness can be estimated as  $K=E a_c$ and $K_{int}=G a_s$ \cite{muser2004}, respectively, where the Young’s modulus $E \sim 0.5-1$ TPa \cite{lee} and intershell shear modulus $G \sim 1$ GPa 
\cite{xlwei} have been measured experimentally. Experimental \cite{servantie} and numerical \cite{lfwang} estimations of the damping coefficient give $\eta \sim 1$ ps$^{-1}$. The mass $m$, can be calculated from the bulk density of carbon 
nanotubes as $m=\rho_V \pi D h a_c$ where $\rho_V=2240$ kg/m$^3$ is the bulk density, $h=0.34$ nm is the wall thickness\cite{lfwang} and $D$ is the diameter. 
As an example of incommensurate DWCNTs consider for instance an armchair tube inserted into a zigzag one, in such a case $a_s - a_c$ is roughly $0.1 a_{C-C}$, where $a_{C-C}=0.142$ nm is the carbon-carbon bond length. For DWCNTs 
with the diameter of the outer shell $D \sim 2.73$ nm pulled with speed $V_0 \sim 1 \mu$m/s through a force probe attached to one its ends \cite{zhang}, $L_{cr}$ is estimated to be $0.5$ m. 
Our theoretical model is thus partially validated by the experiment where the longest DWCNT manipulated was $\sim 9$ mm and is found to display a superlubric behavior. Moreover eq.(\ref{eccola}) suggests that carbon based nanostructures, with a large internal stiffness $K$ due to the $C-C$ bond strength and a small $K_{int}$ due to the weak interplane interaction, are best candidates for further scaling up of superlubricity.

\section{Conclusions}
Summing up, in the framework of a Frenkel-Kontorova-like model, we have addressed the robustness of the superlubricity phenomenon in an edge-driven system at large scales, 
highlighting the mechanisms leading to its failure due to the slider elasticity. The results of the numerical simulations perfectly match the length critical size derived 
from a parameter-free analytical model. By considering different driving and commensurability interface configurations, we have explored the distinctive nature of the transition 
from superlubric to high-friction sliding states which occurs above the critical size, discovering the occurrence of previously undetected multiple dissipative jumps in the friction force 
as a function of the slider length. These  driving-induced commensurate dislocations in the slider are then characterized in relation to their spatial localization and width depending on 
the system parameters. This investigation provides a novel perspective on friction and nanomanipulation experiments and can serve as a theoretical basis for designing nanodevices 
with specific superlow frictional features.

\section{Acknowledgment}
We acknowledge Martin M\"user for insightful comments and remarks. M. U. acknowledges support by the Israel Science Foundation Grant 1316 and by the German-Israeli Project Cooperation Program (DIP). A.V. and A.B. acknowledge support by SINERGIA Project CRSII2 136287/1 from the Swiss National Science Foundation and the ERC Advanced Grant No. 320796-MODPHYSFRICT. This work is also partially supported by the COST Action MP1303 \emph{Understanding and Controlling Nano and Mesoscale Friction}.

\section{Author Contributions}
A.B., M.M., M.U., and A.V. developed the theoretical model, performed
the simulations, and contributed to the numerical data analysis.
All authors discussed the results and contributed to the writing
of the manuscript.

\section{Additional Information}
{\bf Competing financial interests:} The authors declare no competing financial interests.

\end{document}